\documentclass[a4paper,11pt]{article}
\usepackage{pos}
\usepackage{units}

\usepackage{xspace}

\newcommand{\Xmax}{$X_{\rm max}$\xspace} 
\newcommand{\Xmaxmath}{X_{\rm max}}

\newlength{\bibitemsep}
\newlength{\bibparskip}
\let\oldthebibliography\thebibliography
\renewcommand\thebibliography[1]{%
	\oldthebibliography{#1}%
	\setlength{\parskip}{\bibitemsep}%
	\setlength{\itemsep}{\bibparskip}%
}

\title{Constraining the cosmic-ray mass composition by measuring the shower length with SKA}
\ShortTitle{Measurements of the shower length with SKA}

\author*[a,b]{S.~Buitink}
\author[a,b]{A.~Corstanje}
\author[j]{J.~Bhavani}
\author[a]{M.~Desmet}
\author[b, c, d]{H.~Falcke}
\author[e]{B.M.~Hare}

\author[b,d,a]{J.R.~H\"orandel}
\author[f,a]{T.~Huege}
\author[f]{N.~Karasthatis}
\author[a]{G.~K.~Krampah}
\author[a]{P.~Mitra}
\author[a]{K.~Mulrey}

\author[g,h]{A.~Nelles}
\author[b]{K.~Nivedita}
\author[a]{H.~Pandya}
\author[a]{J.~P.~Rachen}

\author[i]{O.~Scholten}
\author[j]{S.~Thoudam}

\author[k]{G.~Trinh}

\author[c]{S.~ter Veen}

\affiliation[a]{Vrije Universiteit Brussel, Astrophysical Institute, Pleinlaan 2, 1050 Brussels, Belgium}

\affiliation[b]{Department of Astrophysics/IMAPP, Radboud University Nijmegen\\
	P.O. Box 9010, 6500 GL Nijmegen, The Netherlands}

\affiliation[c]{Netherlands Institute for Radio Astronomy (ASTRON)\\
	Postbus 2, 7990 AA Dwingeloo, The Netherlands}

\affiliation[d]{Nikhef, Science Park Amsterdam, 1098 XG Amsterdam, The Netherlands}
\affiliation[e]{University of Groningen, Kapteyn Astronomical Institute, Groningen, 9747 AD, Netherlands}
\affiliation[f]{Institut f\"{u}r Astroteilchenphysik, Karlsruhe Institute of Technology (KIT) \\
	P.O. Box 3640, 76021, Karlsruhe, Germany}
\affiliation[g]{DESY, Platanenallee 6, 15738 Zeuthen, Germany}
\affiliation[h]{ECAP, Friedrich-Alexander-University Erlangen-N\"{u}rnberg, 91058 Erlangen, Germany}
\affiliation[i]{Interuniversity Institute for High-Energy, Vrije Universiteit Brussel \\
	Pleinlaan 2, 1050 Brussels, Belgium}
\affiliation[j]{Department of Physics, Khalifa University, P.O.~Box~127788, Abu Dhabi, United Arab Emirates}
\affiliation[k]{Department of Physics, School of Education, Can Tho University Campus II \\
	3/2 Street, Ninh Kieu District, Can Tho City, Vietnam}

\emailAdd{stijn.buitink@vub.be}

\abstract{The current generation of air shower radio arrays has demonstrated that the atmospheric depth of the shower maximum \Xmax can be reconstructed with high accuracy. These experiments are now contributing to mass composition studies in the energy range where a transition from galactic to extragalactic cosmic-ray sources is expected. However, we are still far away from an unambiguous interpretation of the data. Here we propose to use radio measurements to derive a new type of constraint on the mass composition, by reconstructing the shower length $L$.
	
The low-frequency part of the Square Kilometer Array will have an extremely high antenna density of roughly 60.000 antennas within one square kilometer, and is the perfect site for high-resolution studies of air showers. In this contribution, we discuss the impact of being able to reconstruct $L$, and the unique contribution that SKA can make to cosmic-ray science.}

\FullConference{%
  9th International Workshop on Acoustic and Radio EeV Neutrino Detection Activities - ARENA2022\\
  7-10 June 2022\\
  Santiago de Compostela, Spain}

\begin{document}
\maketitle

\section{Introduction}
The low-frequency part of the Square Kilometre Array \cite{Tan:2015} will have roughly $60,000$ antennas within an area of one square kilometer. The antennas are omni-directional and have a large bandwidth of $50-350$~MHz. This makes the SKA a unique site for radio detection of air showers. While the antenna density of LOFAR \cite{LOFAR}  is large within the circular antenna fields, there are also large gaps between those stations. At the SKA, on the other hand, the radio footprint will be homogeneously sampled with thousands of antennas. Such extremely high-resolution measurements offer new reconstruction possibilities that can contribute to cosmic-ray source identification.

The size of the SKA determines the upper limit on the cosmic-ray energy at $\sim 10^{18}$~eV. At lower energies, the strength of the radio signal is the limiting factor. The design and antenna density of SKA allows for a considerable gain in sensitivity by using beamforming. We estimate that the lower energy range will lie around $10^{16}$~eV. This part of the cosmic-ray spectrum, between the knee and ankle, is very complex. It is likely to contain the transition from Galactic to extra-galactic origin. Moreover, it may contain a secondary Galactic component consisting of cosmic-rays that are re-accelerated at the Galactic termination shock or that gain their energy in the strongly magnetized shocks of Wolf-Rayet supernova's \cite{Thoudam}. 

Determining the cosmic-ray mass composition is key to understanding the astrophysics between the knee and ankle. This needs to be measured with high accuracy as some models predict transitions between elements of similar masses. For example, in scenarios featuring helium-rich Wolf-Rayet stars, the transition from Galactic to extra-galactic flux is marked by a change from a helium to proton dominated flux \cite{Thoudam}. The observational challenge is made even larger by the uncertainties in hadronic interaction models. Available models predict different relations between the primary cosmic-ray mass and the average \Xmax of the showers they will generate.

The extremely high antenna density of SKA allows reconstruction of  air showers in unprecedented detail. From Monte Carlo studies \cite{Arthur} it is already clear that traditional radio reconstruction methods will yield a precision of 6 -- 8 g/cm$^2$ on \Xmax and 3\% on primary energy. However, the true potential of cosmic-ray science with SKA can only be understood by developing new analysis ideas that allow us to extract more information from the air showers. Here, we argue that reconstruction of the shower length $L$ is a powerful way to determine the proton fraction of the cosmic-ray flux.

\section{Parametrization of the longitudinal development}
Several parametrizations exist to describe the longitudinal development of air showers. Here we use \cite{Andringa:2011}:
\begin{equation}\label{eq:LRdist}
N(X) = \exp \left(-\frac{X - \Xmaxmath}{RL}\right)\,\left(1 - \frac{R}{L}\left(X - \Xmaxmath\right)\right)^{\frac{1}{R^2}},
\end{equation}
where $N$  is the number of particles in the shower, and $X$ is the traversed depth in the atmosphere in $\unit[]{g/cm^2}$. The parameter $L$ scales with the width of the profile and is a measure for the length of the shower, while $R$ is a measure for the asymmetry in the profile shape before and after the shower maximum.

The Pierre Auger Observatory has performed a measurement of the average $L$ and $R$ for a large set of showers based on the fluorescence technique \cite{Auger_RL}. There are now several indications that it is possible to use radio observations to reconstruct $L$ for individual showers. 

\begin{figure}
	\centering
	\includegraphics[width=0.7\textwidth]{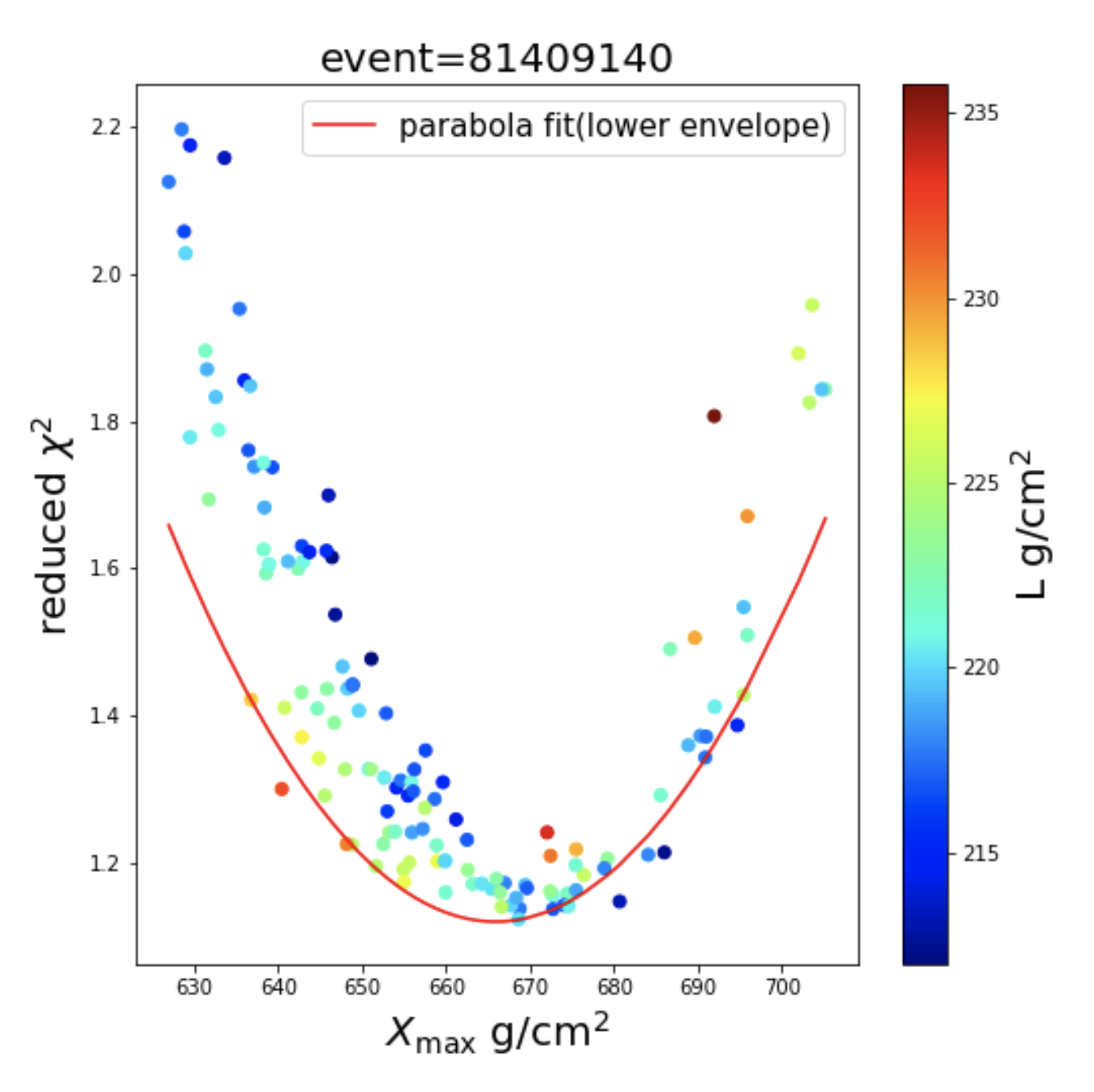}
	\caption{Each dot represents a simulated shower that is fitted to LOFAR data. The reduced $\chi^2$ of the fit is plotted against \Xmax of the simulated shower. In the LOFAR shower reconstruction such plots are used to reconstruct \Xmax by fitting a lower envelope parabola. For this event, a strong scatter is observed. The color coding demonstrates that this scatter is primarily the result of different values for $L$.} 	\label{fig:LOFAR_L}
\end{figure}

The first indication comes from LOFAR data. To reconstruct \Xmax, the radio emission profiles of a large set of simulated showers are fitted to the data \cite{Buitink14}. Figure \ref{fig:LOFAR_L} shows that the reduced $\chi^2$ of the fit critically depends on both \Xmax and $L$. Reconstructing both parameters would require a much larger set of simulations, which are very computationally expensive. Future analysis possibilities will critically depend on the development of faster simulation techniques, such as the template synthesis method \cite{template_synthesis}.   

The second indication comes from CORSIKA/CoREAS simulations for the SKA \cite{CORSIKA, CoREAS}. First results show that it is indeed possible to simultaneously reconstruct \Xmax and additional information on the longitudinal development of the shower \cite{Arthur}. In particular, it has been shown that the most sensitive parameter after \Xmax is a linear combination of $L$ and $R$ . It should be noted that this result is from a first exploration of the possibilities and there is ample room for improvement: the optimal reconstruction techniques have yet to be determined. 

In the remainder of  this contribution, we will investigate the impact of measurements of $L$ on constraining the mass composition. We assume that SKA will be able to reconstruct $L$ with a resolution of  $\sim 10$~g/cm$^2$ \cite{Arthur}.

\section{CONEX simulations}
\begin{figure}
	\centering
	\includegraphics[width=0.9\textwidth]{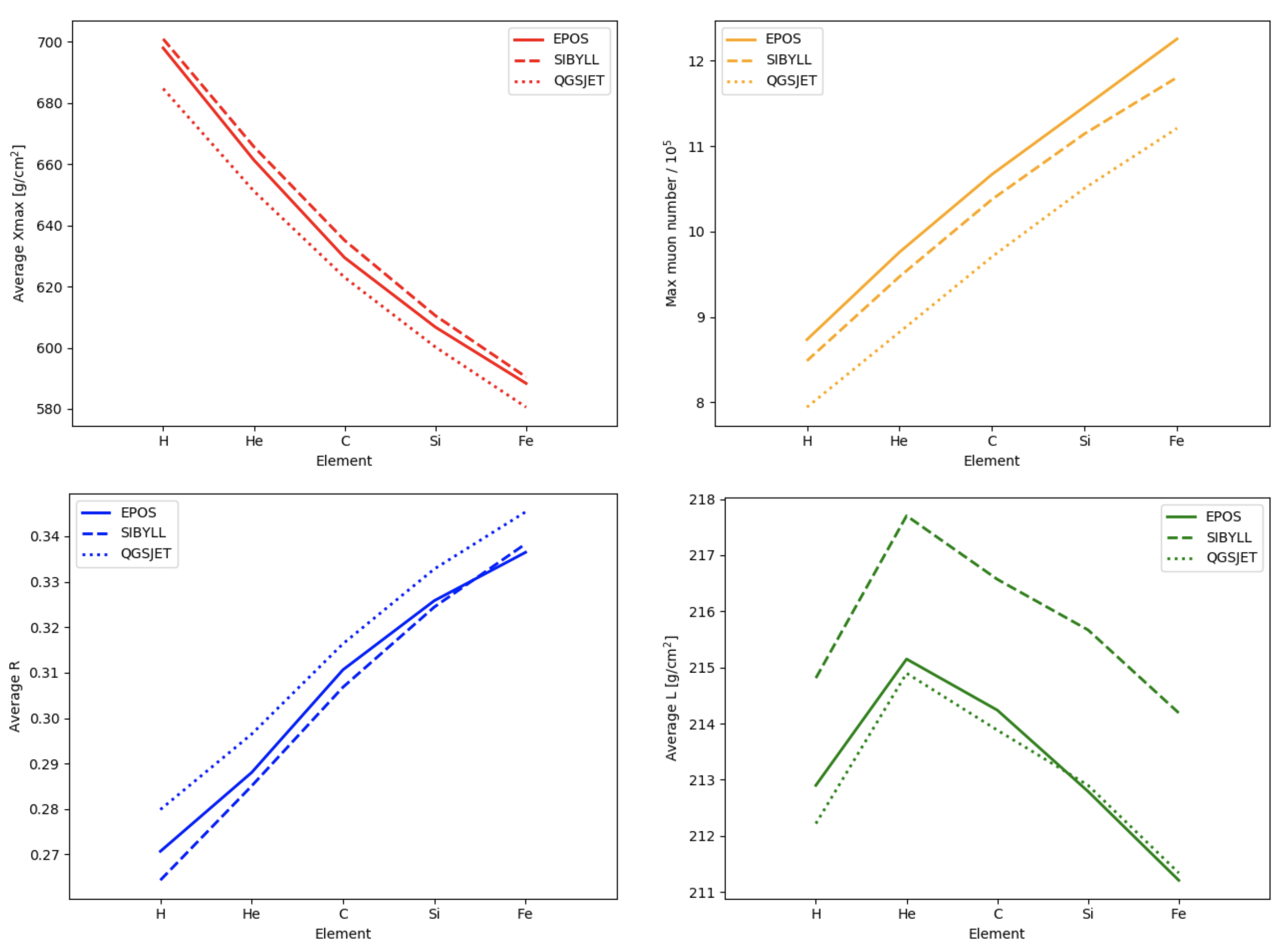}
	\caption{Results from CONEX simulations. Average \Xmax, $R$, and $L$ as a function of primary element for different hadronic interaction models.}
	\label{fig:trends}
\end{figure}

\begin{figure}
	\centering
	\includegraphics[width=0.9\textwidth]{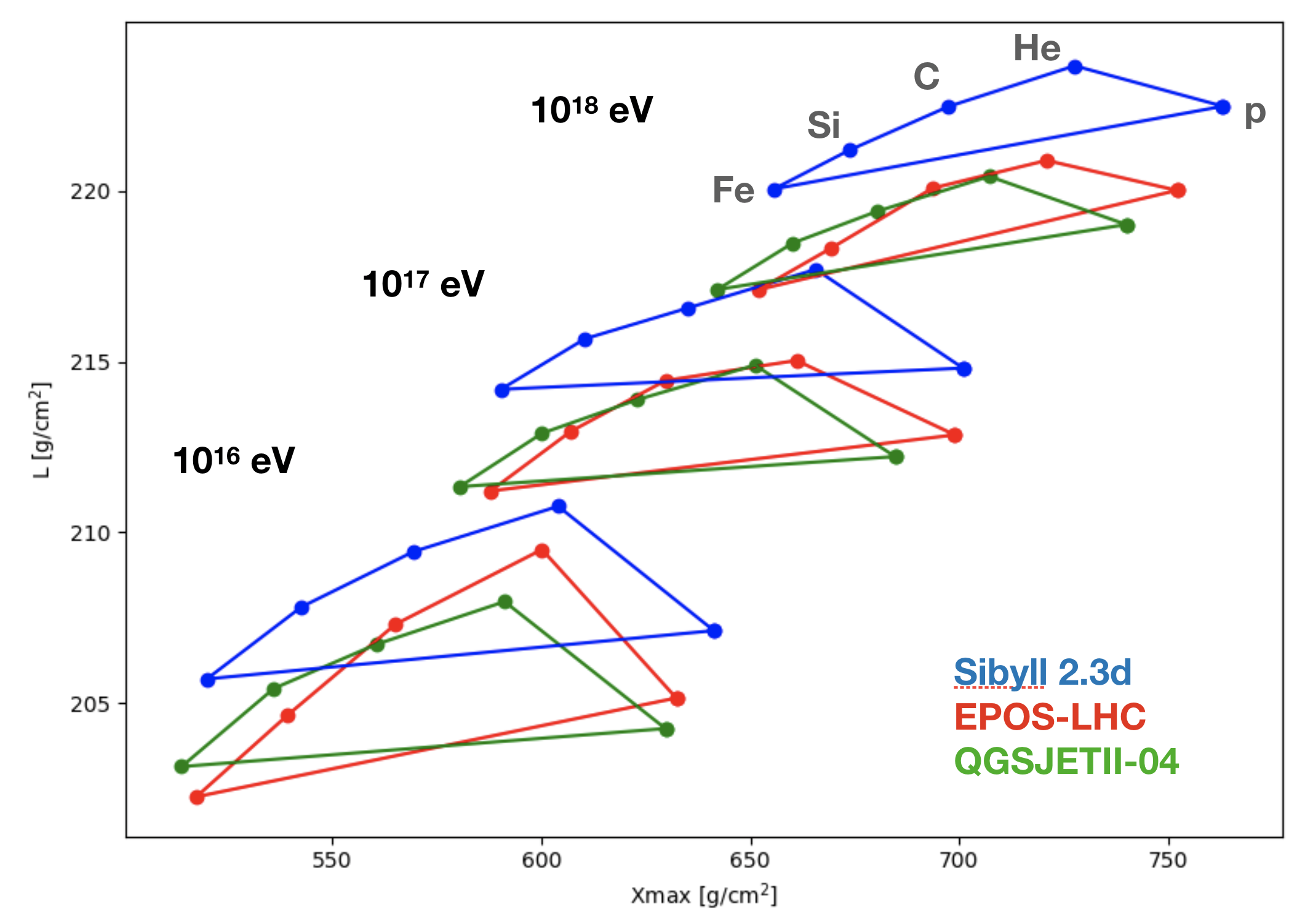}
	\caption{Average $L$ versus average \Xmax for different energies and hadronic interaction models. In each case, the elements form a triangle with protons on one corner and all other elements on the opposite side.}
	\label{fig:triangles}
\end{figure}

We have set up CONEX simulations at three energies (10$^{16}$~eV, 10$^{17}$~eV, 10$^{18}$~eV) for three hadronic interaction models (EPOS-LHC, QGSJETII-04, and Sibyll 2.3d), and five elements (proton, helium, carbon, silicon, iron). For each combination of parameters we have generated 2500 showers.  Their longitudinal profiles are fitted with Eqn.~\ref{eq:LRdist} to obtain \Xmax, $L$, and $R$.

In most cases the parametrization fits the longitudinal profile very well, but there are exceptions. Occasionally, showers have a double-bump structure. These structures occur more often for proton and helium showers at the lowest energies. Even then, however, the fraction of showers with a bad fit is below 0.5\%. 

Figure~\ref{fig:trends} shows the average values of \Xmax, $R$, $L$, and the number of muons at shower maximum, for different elements and hadronic interaction models. Whereas most parameters feature a monotonous trend from light to heavy elements, $L$ increases sharply from proton to helium and then drops steadily for further increasing primary mass. A combined measurement of $L$ and any of the other parameters allows us to isolate protons from all other elements.

Figure~\ref{fig:triangles} shows the average \Xmax and $L$ for all generated elements, energies, and hadronic interaction models. For each combination of energy and interaction model, the elements form a triangle with protons on one corner, and all other elements on the opposite side. For some triangles, this side is curved, but the protons never lie on the same curve as the other elements. 

The average $L$ predicted by the three hadronic interaction models diverges towards higher energies. In particular, Sibyll 2.3d produces higher values for $L$ than the other two models. The real cosmic-ray flux is some mixture of elements and would therefore have an average \Xmax and $L$ that falls somewhere inside the triangle. Observations of \Xmax and $L$ can thus be used to test the validity of models, even with incomplete knowledge of the mass composition. SKA can in this way provide a unique measurement to probe the hadronic processes in the shower.

\section{Robust measurements of L}
\begin{figure}
	\centering
	\includegraphics[width=0.7\textwidth]{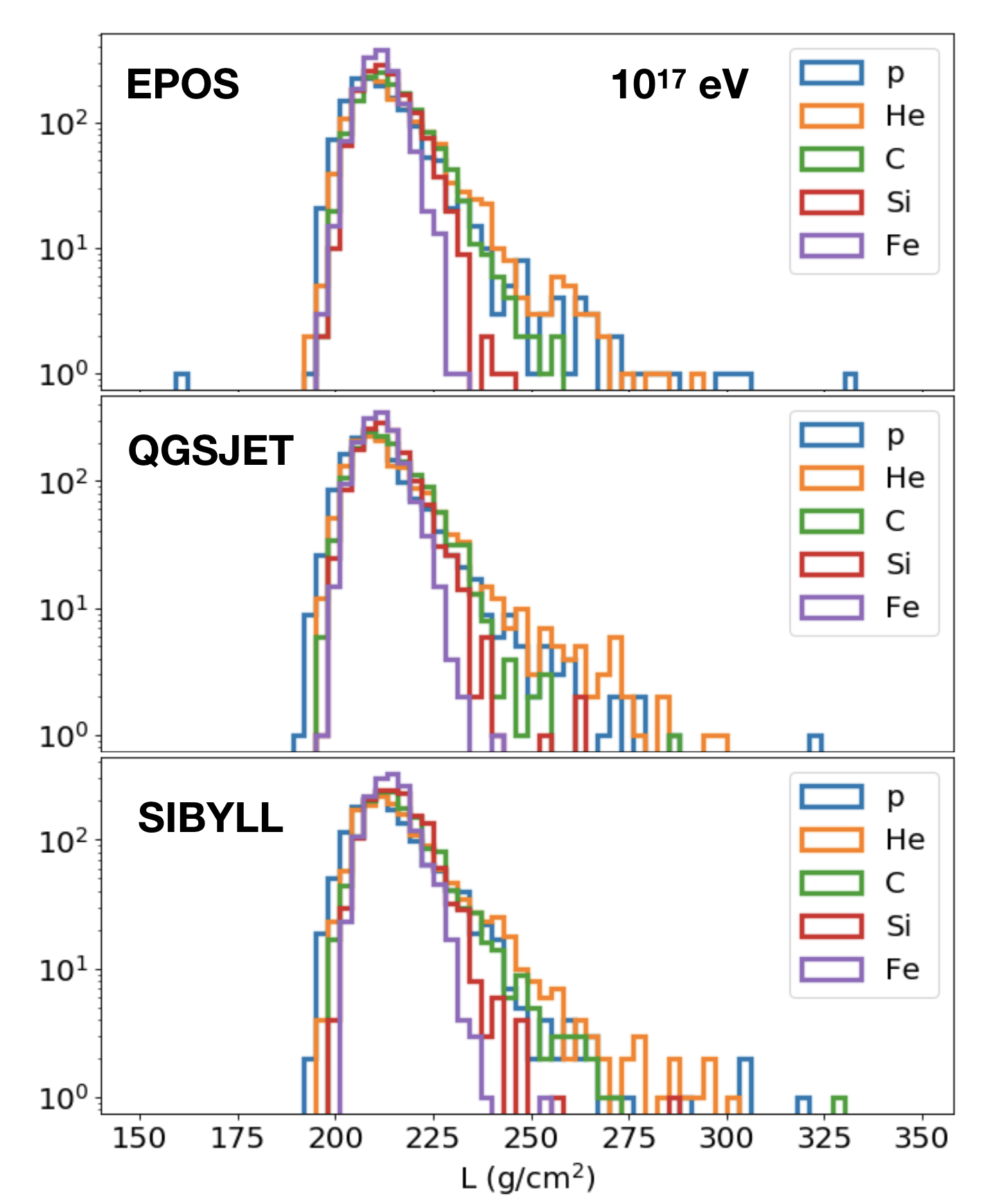}
	\caption{Histograms of fitted $L$ for all elements and interaction models. The shape of the high-$L$ tail of the distribution is a more robust mass-dependent parameter than the average $L$.}
	\label{fig:histo}
\end{figure}

While $L$ can provide important information about the mass of the primary, it remains to be proven that it can be reconstructed robustly enough to use it in a mass composition analysis. The differences in average $L$ between different elements are of  the order of  a few $\unit[]{g/cm^2}$. We do not have a solid prediction yet for the expected systematic uncertainty on reconstructing $L$ with SKA, but the level of accuracy needed seems very challenging. However, there are more robust features in the distribution of $L$.

Figure \ref{fig:histo} shows histograms of $L$ for $10^{17}$~eV showers of different primary masses. There is good agreement between the three hadronic interaction models. While the peaks of the distributions of all elements are close together, the high-$L$ tails are very different. Determining the width of the distribution, or the shape of the tail is therefore a much more robust observable, that can tolerate higher systematic uncertainties. For illustration purposes, we will here adopt a simple test parameter: the fraction of showers with $L > 225$~g/cm$^{2}$.

\section{Determining the proton fraction}

\begin{figure}
	\centering
	\includegraphics[width=0.44\textwidth]{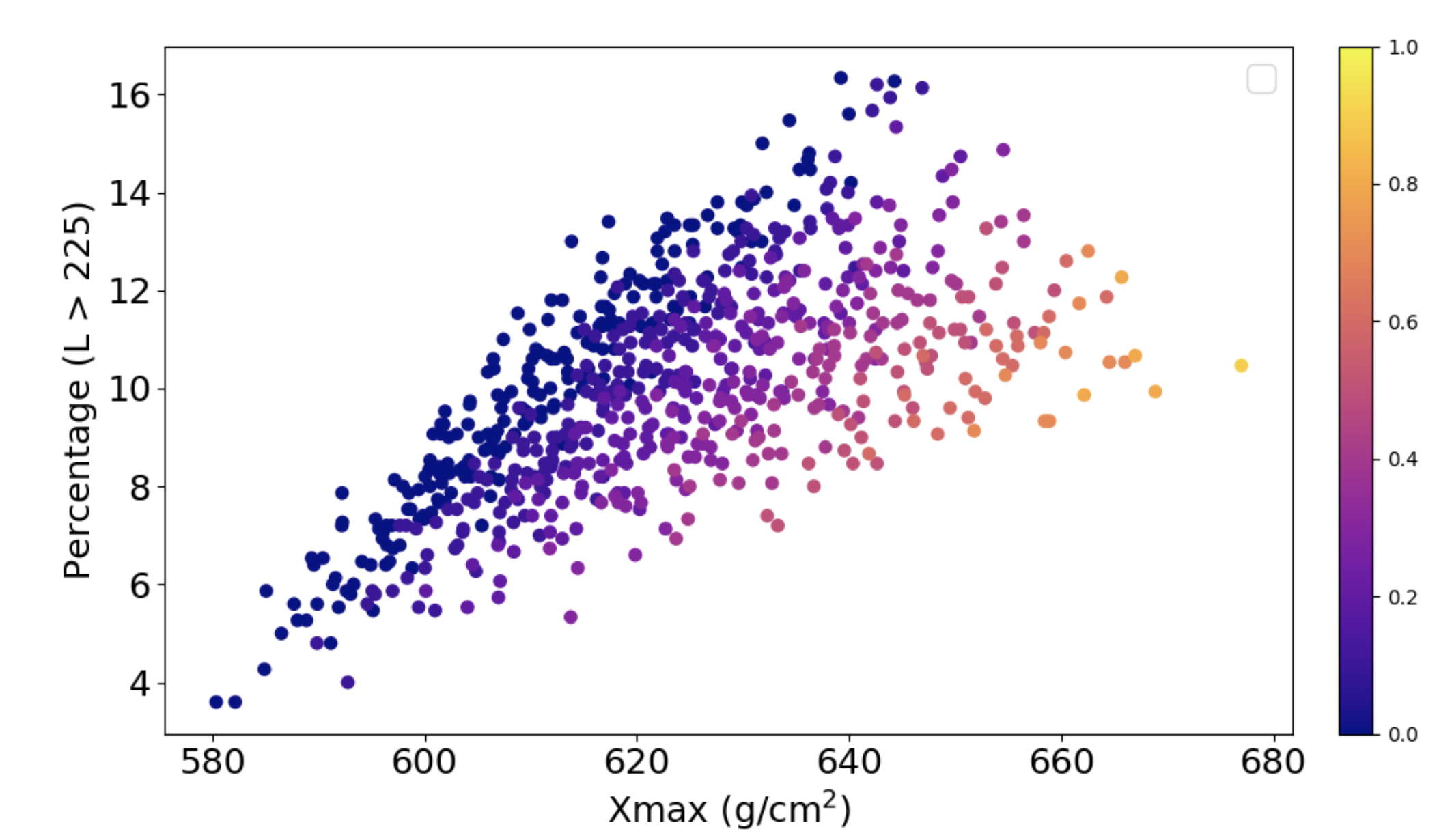}
	\includegraphics[width=0.54\textwidth]{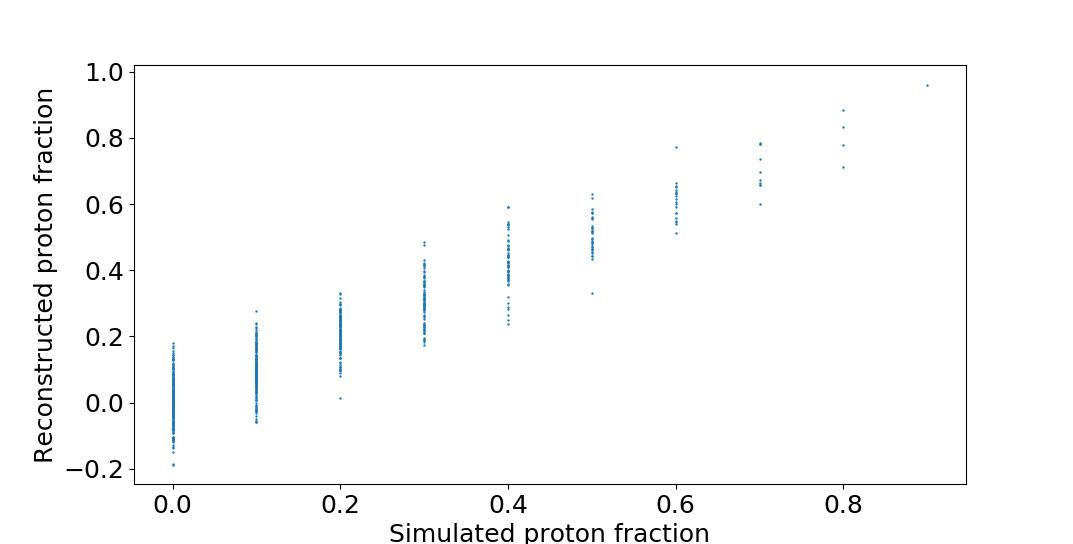}
	\caption{Left: Each dot represents a set of 1500 showers of a random mix of elements. The color indicates the proton fraction. The dot is plotted at the position corresponding to the average \Xmax of the sample and the percentage of showers with $L>225$~g/cm$^2$.  Right: Reconstructed proton fraction versus simulated proton fraction for all random shower sets, obtained with toy analysis.}
	\label{fig:ProtFrac}
\end{figure}
To test the impact of reconstructing $L$ on the mass composition analysis we use the CONEX shower library to generate random astrophysical models. Each model consists of the five elements in our set, but the mixing ratios are different. For each model, we randomly select 1500 conex showers, and add random measurement errors to the \Xmax and $L$ values of each individual shower. For both parameters we add a random error following a Gaussian distribution  of $\sigma=10$~g/cm$^{2}$, which is a reasonable choice for SKA\cite{Arthur}. 

For each model, the average \Xmax and the fraction of $L > 225$~g/cm$^{2}$ is plotted in the left panel of Fig.~\ref{fig:ProtFrac}. The color coding indicates the proton fraction in the model. We again recognize the triangle shape. Every model with 0\% protons will fall somewhere on the left side of the triangle. The position on this (slightly curved) line is determined by the actual mixing ratios of helium, carbon, silicon, and iron. Models with larger proton fraction will fall on a similar curve, but shifted towards the right. This is seen by the clean trend in color in the Figure. 

We can now set up a simple toy analysis to reconstruct the proton fraction. The analysis is based on the simulated positions of a pure proton, helium, and iron sample. First, a line is reconstructed that passes through the pure iron and helium points. Next, we determine the distance $D$ of the pure proton point to this line. Finally, we determine the distance $d$ to the line for each of the generated mock data sets. The reconstructed proton fraction for this data set is then given by $d/D$.  From Fig.~\ref{fig:triangles} it is clear that intermediate mass elements can be on the left side of the line connecting iron and helium. These are assigned negative values for $d$, resulting in a  negative reconstructed proton fraction.

The results are shown in the right panel of Fig.~\ref{fig:ProtFrac}.  The proton fraction is retrieved with a precision of $\sim 10$\%. Since this analysis is completely un-optimized it is likely that even better results could be achieved. While this result is obtained with a 5-element model, the precision is likely to be similar when including more elements, since all elements other than proton lie on the same curve.

\section{Conclusion}
Radio observation of air showers with the SKA will provide unprecedented detail and precision. Existing reconstruction techniques will not use the observatory to its full potential. Monte Carlo simulations have demonstrated that it will be possible to reconstruct the shower length $L$ for individual showers. Unlike most other shower observables, $L$ does not scale monotonously with the primary mass, but is largest for helium. This feature can be used to design new analysis techniques that separate the proton showers from all other mass components. This is invaluable for understanding the cosmic-ray sources in the Galactic-to-extra-galactic transition region. In addition, measurements of $L$ will provide a unique measurement that can put constraints on hadronic interaction models.   

\section*{Acknowledgements}
TNGT acknowledges funding from Vietnam National Foundation for Science and Technology Development (NAFOSTED) under grant number 103.01-2019.378. ST acknowledges funding from the Khalifa University Startup grant, project code 8474000237-FSU-2020-13.

\end{document}